\documentclass[manuscript]{aastex7}

\usepackage[version=4]{mhchem}
\usepackage{CJK}

\begin{document}

\begin{CJK*}{UTF8}{gbsn}
  
\title{Retention of surface water on tidally locked rocky planets in the Venus zone around M dwarfs}

\author[orcid=0009-0004-4797-8791, gname='Yueyun', sname='Ouyang']{Yueyun Ouyang}
\affiliation{Department of Atmospheric and Oceanic Sciences, School of Physics, Peking University, Beijing 100871, China}
\email{2301110312@stu.pku.edu.cn}

\author[orcid=0000-0001-7758-4110, gname='Feng', sname='Ding']{Feng Ding (丁峰)} 
\affiliation{Department of Atmospheric and Oceanic Sciences, School of Physics, Peking University, Beijing 100871, China}
\email[show]{fengding@pku.edu.cn}

\author[orcid=0000-0001-6031-2485, gname='Jun', sname='Yang']{Jun Yang (杨军)} 
\affiliation{Department of Atmospheric and Oceanic Sciences, School of Physics, Peking University, Beijing 100871, China}
\email{junyang@pku.edu.cn}

\begin{abstract}

Terrestrial planets within the Venus zone surrounding M dwarf stars can retain surface ice caps on the perpetual dark side if atmospheric heat transport is inefficient,  {as suggested by previous global climate simulations \citep[e.g.,][]{leconte2013}.} This condition is  {proposed} to play a role in the potential regional habitability of these planets. However, the amount of surface ice may be limited by considering the water condensed from the steam atmosphere in a runaway greenhouse state, and the physical mechanism for triggering the condensation process is not clear. Here, we use a two-column moist radiative-convective-subsiding model to investigate the water condensation process on tidally locked planets from the runaway greenhouse state. We find that the water condensation process is characterized by two distinct equilibrium states under the same  {incoming stellar flux}. The initiation of condensation corresponds to a warm, unstable state exhibiting positive Planck feedback, whereas the termination phase corresponds to a cold, stable state exhibiting negative Planck feedback. We further show that the surface water mass in the collapsed state  {decreases with the} incoming stellar flux, background surface pressure, and optical thickness of non-condensible greenhouse gases, with a global equivalent depth of less than $\sim 20$ cm. Our two-column approach provides a straightforward way to understand the water evolution on Venus zone planets around M dwarfs.

\end{abstract}



\section{Introduction} 

The transit method, an effective way to discover planets beyond our solar system, is intrinsically biased towards planets with shorter orbital periods \citep{winn2010exop.book.transit, Kane_2008transit}, which has led to more terrestrial planets discovered in the Venus zone than those in the habitable zone \citep{Kane_2014venuszone, Ostberg_2023venuszone,peterson2023temperate}. The Venus zone was first proposed by \cite{Kane_2014venuszone} and was defined as a circumstellar zone around the host star, with the inner and outer boundary determined by the `Cosmic Shoreline' \citep{Zahnle_2017cosmicshoreline} and the runaway greenhouse boundary \citep{kasting1993habitable,Yang_2013cloud}, respectively. Therefore, terrestrial planets in the Venus zone can potentially have an atmosphere against escape processes but cannot maintain a global surface ocean as Earth-like planets.

\begin{figure}[ht!]
\centering
\includegraphics[width=0.8\textwidth]{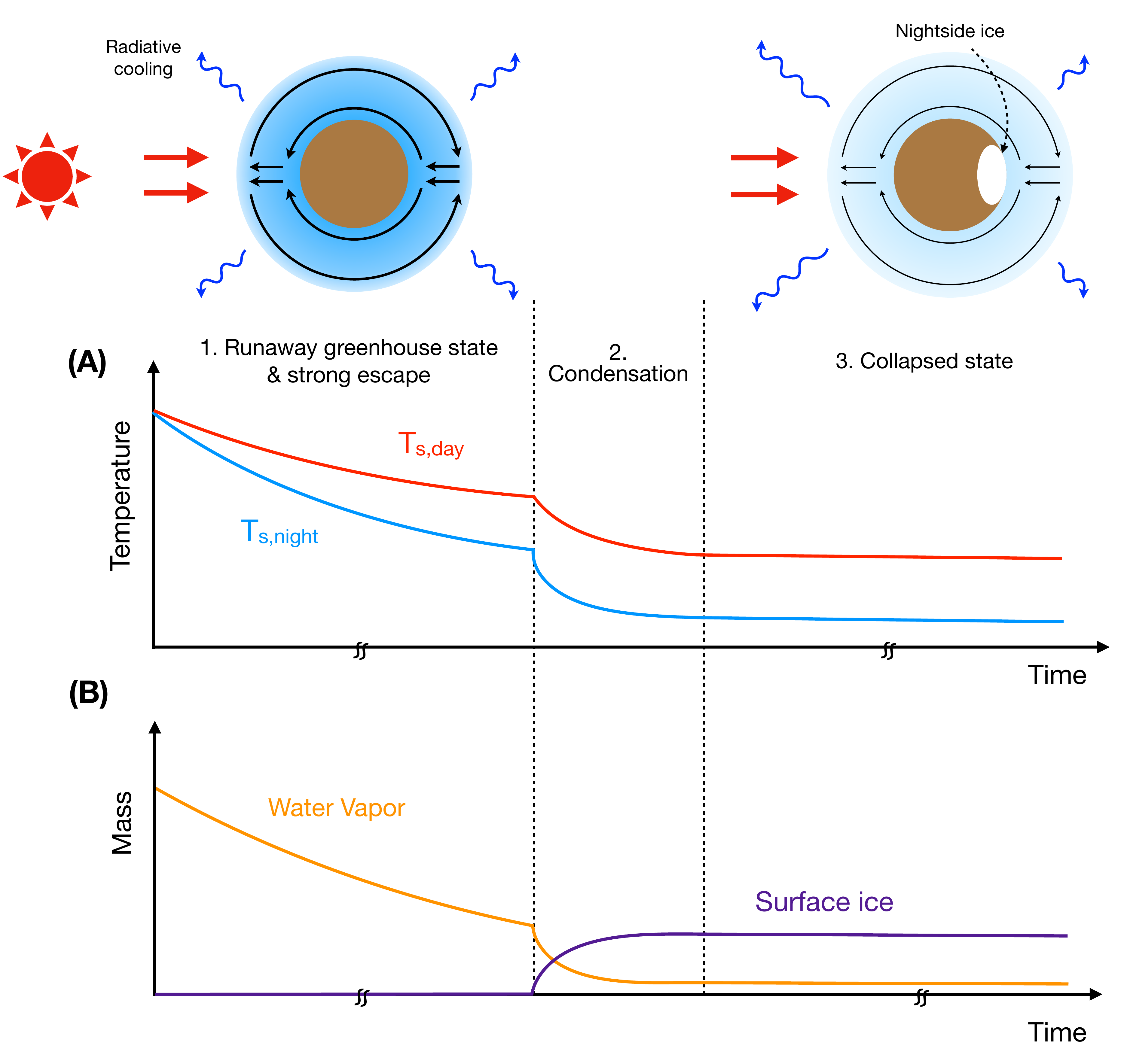}
\caption{Schematic diagram illustrating  {a} possible climate evolution of tidally locked planets in the Venus zone. Stage 1: primordial steam-dominated atmosphere is gradually lost by intense stellar radiation and atmospheric heat transport is weakened. Stage 2: water began to condense from the atmosphere to the surface. Stage 3: the planet is trapped in the collapsed state. Red and blue curves in panel(A) represent the evolution of surface temperature on the dayside and nightside, respectively. Orange and purple curves in panel(B) represent the evolution of atmospheric and surface water mass, respectively.}
\label{fig:evolution}
\end{figure}

Recent three-dimensional (3D) global climate modelings suggest that, as the primordial steam-dominated atmosphere on terrestrial planets in the Venus zone is gradually lost by intense stellar radiation, atmospheric  {horizontal} heat transport will be weakened  {\citep[e.g., ][]{leconte2013,koll2016}}. These planets may evolve from the runaway greenhouse state into the collapsed regime (see Figure~\ref{fig:evolution} for details), in which water began to condense from the atmosphere and be trapped at the surface in permanent cold traps on the night side  {of synchronously rotating planets} \citep{leconte2013, Ding_2021lastsaturation,ding2022prospects} or at the poles  {of asynchronously rotating planets} \citep{abe2011habitable,Kodama_2018dryplanet}. Surface water could be in the form of liquid or ice, depending on the surface temperature of the permanent cold traps. In addition, long-lived liquid water has also been proposed to possibly be present near the edge or at the bottom of the ice caps due to basal melting and gravity-driven glacier flow \citep{leconte2013,ojha2022liquid,wandel2023habitability}. The possible presence of liquid water on close-in terrestrial exoplanets in the collapsed regime raises an intriguing question about the regional habitability of this type of planets \citep{lobo2023terminator}.

However, fundamental questions are not fully resolved, especially the underlying physical mechanism driving the water transition from the runaway greenhouse atmosphere to the collapsed state and key factors influencing the amount of surface water in the collapsed state, which are crucial for understanding the water evolution and possible regional habitability of Venus zone planets. In addition, the runaway greenhouse effect, widely applied to ocean-covered planets as a criterion for the inner edge of the habitable zone \citep{kasting1993habitable,Yang_2013cloud}, requires further investigation into its role in climate transitions on planets in the Venus zone with limited surface water.

In this Letter, we  {will focus on tidally locked terrestrial planets in the Venus zone around low-mass stars. The emergence of the collapsed state on this type of planets was first studied by 3D global climate simulations in \cite{leconte2013}}. Here, we introduce a two-column framework to address the underlying physical mechanism. The novelty of our two-column framework is its ability to identify unstable equilibrium states of the moist climate system and track their dynamical behavior, which is technically difficult to achieve in 3D general circulation models (GCMs) relying on numerical iterations to find equilibrium states. We will show that the formation of surface water is triggered by a strong positive Planck feedback near the unstable equilibrium state and is intrinsically distinct from the runaway greenhouse effect at the inner boundary of the habitable zone. 

\section{Method} \label{sec:method}

Venus zone planets around low-mass stars are predicted to be in spin-orbit resonance, most likely to be in synchronous rotation, due to their proximity to host stars and strong tidal dissipation \citep{goldreich1966spinorbit,heller2011tidal,barnes2017tidal}. \cite{koll2016} proposed that the 3D atmospheric circulation on synchronous rotating planets can be simplified in a two-column framework with dayside circulation represented by convective motion in one column and nightside circulation represented by subsiding motion in the other, and the strength of global circulation is constrained by the efficiency of the atmospheric heat engine. As a result, the thermal structures in the two columns can be derived from the incoming stellar flux $S_\ast$ and the atmospheric infrared optical thickness $\tau$ \citep{koll2016, Auclair_Desrotour_2020twocolomn}. 

Based on  {the dry radiative-convective-subsiding (RCS) model developed by \cite{koll2016}}, we place a new constraint in the model to simulate the collapsed state of Venus zone planets and separate the infrared optical thickness into components associated with dry greenhouse gases and water vapor ($\tau_{LW} = \tau_{dry} + \tau_{wv}$). GCM simulations indicate that in the collapsed state, water vapor is well mixed in the atmosphere and is in phase equilibrium with the nightside surface water \citep{ding2020stabilization,Ding_2021lastsaturation}. Therefore, the infrared optical thickness of water vapor $\tau_{wv}$ is connected with the nightside surface temperature $T_{s,n}$ by the Clausius-Clapeyron relation
\begin{equation} \label{eq:tauwv}
\tau_{wv} = \frac{q_{sat}(T_{s,n}, p_s)}{g \bar{\mu}}{\int^{p_s}_0\overline{\kappa}_{wv}(p) \text{d}p}, 
\end{equation}
where $q_{sat}, p_s, g, \bar{\mu}$ are the saturation mass fraction of water vapor at the nightside surface, surface pressure, surface gravity, and the mean cosine of infrared emission angle, respectively  {(see Appendix~\ref{sec:calc_tau} for details)}. $\bar{\kappa}_{wv}$ is the Rosseland mean infrared absorption cross section of water vapor calculated by the line-by-line radiation code PyRADS\footnote{\url{https://github.com/danielkoll/PyRADS}}, which has been used to study the greenhouse effect of water vapor on Earth and other exoplanets \citep{koll2018,Koll_2019}.
With our new constraint (Eq.~\ref{eq:tauwv}), semi-analytical solutions representing the collapsed state with surface water trapped on the nightside surface can be directly derived given the incoming stellar flux $S_\ast$ and the dry infrared optical thickness $\tau_{dry}$. 

Our two-column moist RCS (MRCS) model can approximately reproduce water condensation in an Earth-like atmosphere on exoplanet Gl~581c  {with varying background surface pressures} simulated by LMD GCM  {(see Appendix~\ref{sec:reproduce} for details)}. To better compare our work with complex GCM simulations, we will show results using Gl~581c's mass and radius in this paper. Similarly to \cite{leconte2013}, the non-condensible background atmosphere is assumed to be composed of \ce{N2} and  {\ce{CO2}} gases with  {background surface pressure of $p_{dry}$ and} optical thickness of $\tau_{dry}$,  {where the background surface pressure $p_{dry} = p_{s,\ce{N2}} + p_{s,\ce{CO2}}$, and $p_{s,\ce{N2}}$ and $p_{s,\ce{CO2}}$ are the surface partial pressure of \ce{N2} and \ce{CO2}, respectively. Then the molar concentration of \ce{CO2} is calculated as $f_{\ce{CO2}} = p_{s,\ce{CO2}} / p_{s}$. The mean optical thickness of \ce{CO2} is evaluated by Eq.~4 in \cite{koll2019secondary}.} The values of the parameters used in this work are summarized in Table \ref{tab:params}. 
 
\begin{table}[]
    \centering
    \begin{tabular}{cc}
    \hline
    \hline
    Parameters       & Values \\
    \hline
    Incoming stellar flux ($S_\ast / S_\oplus$) & 2.35-8.08 \\ 
    Planetary radius ($R_p/R_\oplus$) & 1.85 \\
    Surface gravity ($g$) & 18.4 m s$^{-2}$\\
    Surface albedo ($\alpha$) & 0.3 \\
    Background surface pressure ($p_{dry}$) & 0.5-2 bar\\
    molar concentration of \ce{CO2} ($f_{\ce{CO2}}$)  & 188-37600 ppmv\\
    Mean cosine of infrared emission angle ($\bar{\mu}$) & 0.6\\
    \hline
    \end{tabular}
    \caption{Parameters used in our simulations. The planetary mass and radius are the same as those of Gl~581c. The incoming stellar flux of Gl~581c is 2.46$S_\oplus$, where the solar constant of Earth $S_\oplus = 1361$ W m$^{-2}$.}
    \label{tab:params}
\end{table}

\section{Results} 

\subsection{Mechanisms driving the nightside condensation}\label{sec:mechanism}

\begin{figure}[!ht]
\centering
\includegraphics[width=0.6\textwidth]{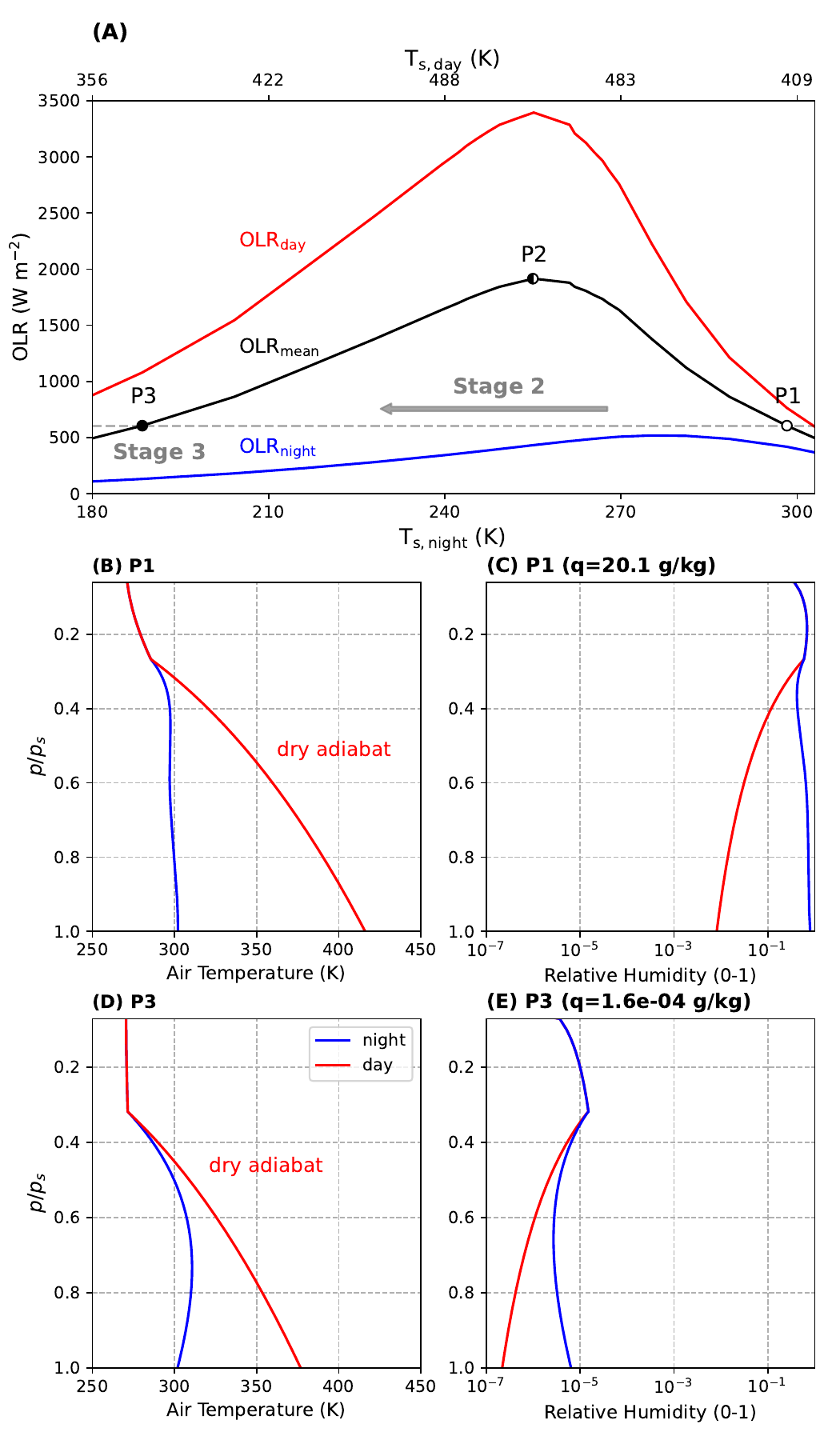}
\vspace{-10pt}
\caption{(A) Relation between the global mean outgoing longwave radiation (OLR) and the surface temperature of the nightside (lower $x$-axis) when $p_{dry}=1$ bar and $f_{\ce{CO2}}=376$ ppmv. The upper $x$-axis shows the corresponding surface temperature of the dayside, and the scale is not linear. States P$_1$ and P$_3$ correspond to equilibrium states with the same OLR of 587.2 W m$^{-2}$, and point P$_2$ marks the maximum OLR value emitted by the planet. Red and blue curves further show the dayside and nightside OLR as a function of the nightside surface temperature. (B,D) Vertical temperature profiles in the dayside (red) and nightside (blue) columns of states \ce{P1} (B) and \ce{P3} (D). (C,E) Vertical profiles of relative humidity in the dayside (red) and nightside (blue) columns that measures the degree of water vapor saturation in the air column of states \ce{P1} (C) and \ce{P3} (E). The title of each panel gives the value of uniform specific humidity in the atmosphere.}\label{fig:OLR_Tn}
\end{figure}

The relation between the global mean outgoing longwave radiation (OLR) and $T_{s,n}$ is given by the black curve in Figure~\ref{fig:OLR_Tn}A when $p_{dry}=1$ bar and $f_{\ce{CO2}}=376$ ppmv. The global mean OLR values are much higher than the Simpson-Nakajima limit \citep{simpson1928runaway,nakajima1992runaway} or the troposphere asymptotic limit \citep{kasting1993habitable,goldblatt2012runaway} used to define the inner edge of the habitable zone. Furthermore, the global mean OLR varies non-monotonically with $T_{s,n}$, while for ocean planets  {with 1 bar \ce{N2} in the atmosphere} near the inner edge of the habitable zone, their OLRs roughly increase with surface temperatures in both 1D and 3D GCM models  {with a small overshoot} before reaching the Simpson-Nakajima limit \citep{kasting1993habitable,Koll_2019,yang2019oceaninneredge}.

This non-monotonic behavior of the global mean OLR is the key mechanism driving the transition from the runaway greenhouse state to the collapsed state, by providing two equilibrium states under the same stellar flux. For example, the points P$_1$ and P$_3$ in Figure~\ref{fig:OLR_Tn}A correspond to equilibrium states under $S_\ast = 2.46 S_\oplus$, where the solar constant of Earth $S_\oplus = 1361$ W m$^{-2}$. \added{The onset of water vapor condensation on the night surface (boundary between stages 1 and 2 in Figure~\ref{fig:evolution}) and the final equilibrium state of the collapsed regime (stage 3 in Figure~\ref{fig:evolution}) are connected to the climate stability at \ce{P1} and \ce{P3}. }
The global mean OLR declines with $T_{s,n}$ around  \ce{P1}, indicating that a cooler climate \added{with lower $T_{s,n}$ value} emits even more energy into space. This positive Planck feedback makes \ce{P1} an unstable state. As long as the nightside surface temperature $T_{s,n}$ is warmer than that of P$_1$, the climate remains in the runaway greenhouse state under $S_\ast = 2.46 S_\oplus$ (stage 1 in Figure~\ref{fig:evolution}). Once $T_{s,n}$ reaches the value of P$_1$ due to atmospheric escape \added{and the associated decline of atmospheric infrared opacity and surface temperature}, atmospheric water vapor starts to condense on the night surface (boundary between stages 1 and 2 in Figure~\ref{fig:evolution}). The condensation of surface water will persist until $T_{s,n}$ reaches the value of P$_3$, where the Planck feedback is negative and a cooler climate emits less energy into space. Therefore,   \ce{P3} is stabilized by the negative Planck feedback and represents the equilibrium state of the collapsed regime (stage 3 in Figure~\ref{fig:evolution}).

We also plot the vertical profiles of air temperatures in both the day and night columns to investigate why the distinct states   \ce{P1} and  \ce{P3} are associated with the identical OLR value. \added{The surface temperature of \ce{P1} is higher than that of \ce{P3} (Figure~\ref{fig:OLR_Tn}A), which is supposed to induce a higher OLR. However, the state \ce{P1} has a higher water vapor content (Figure~\ref{fig:OLR_Tn}C and E) to absorb most of the thermal radiative flux from the surface, drastically reducing its OLR.}
The net effect is that the two factors cancel each other out. This cancelation effect can be clearly seen from the thermal emission of the day column, where the air is hot and dry with the vertical temperature profile following the dry adiabat in the troposphere (Figure~\ref{fig:OLR_Tn}B and D). 
Under an all-troposphere dry atmosphere assumption, the OLR from the day column OLR$_{day}$ can be directly separated by contributions from a thermal factor associated with the surface temperature and a opacity factor $f(\tau_{LW})$ following Chapter 4.3.2 of \cite{pierrehumbert2010ppcbook}:
\begin{equation}
    \mathrm{OLR}_{day} \approx \sigma T_{s,day}^4 f(\tau_{LW}), \quad f(\tau_{LW}) = \exp(-\tau_{LW}) + \tau_{LW}^{-2R/c_p} \gamma(2R/c_p+1, \tau_{LW}), 
\end{equation}
where $ T_{s,day}$ is the surface temperature of the day column, $\gamma$ is the lower incomplete gamma function and $R/c_p = 2/7$ for the \ce{N2}-dominated atmosphere in our simulation. The opacity factor $f(\tau_{LW})$ is a decreasing function of $\tau_{LW}$. Although the surface temperature of the dayside of \ce{P1} is $\sim 40$ K warmer than that of \ce{P3}, the water vapor concentration of \ce{P1} is five orders of magnitude higher than that of \ce{P3}. Then the opacity factor $f(\tau_{LW})$ plays a dominant role in determining the thermal emission to space and exceeds the thermal factor, making OLR$_{day}$ of \ce{P1} $\sim 300 \mathrm{W m^{-2}}$ less than that of \ce{P3}.

The physical explanation given above shares many similarities with the Komabayashi-Ingersoll model of the runaway greenhouse effect \citep{komabayashi1967runaway,ingersoll1969runaway}. The 1D Komabayashi-Ingersoll model proposes that warming the Earth's stratosphere including the tropopause can both increase the thermal factor and decrease the opacity factor for the OLR, and therefore leads to a non-monotonic behavior of OLR. Although the Komabayashi-Ingersoll model is practically impossible to achieve on Earth because  {it neglects processes in the moist troposphere \citep{goldblatt2012runaway} }, our simulation suggests that it can be applied to planets in the Venus zone. 
 {The water vapor is indeed well mixed in the atmosphere and in phase equilibrium with the nightside surface in our MRCS model, in analogy to stratospheric water vapor in phase equilibrium with the tropopause in the Komabayashi-Ingersoll model. Our MRCS model can be considered as a modified Komabayashi-Ingersoll model in a two-column framework.}

\subsection{Implication for surface water content} \label{sec:factors}

Nightside condensation from the state P$_1$ to P$_3$ discussed in Section~\ref{sec:mechanism} is a fast process compared to atmospheric escapes. Previous GCM simulations suggested that it only takes a few years to reach the stable equilibrium state P$_1$ (Figure~7 in \cite{leconte2013}), during which the loss of water due to atmospheric escape is negligible. Then the mass of surface water on the nightside $m_{water}$ can be estimated by the saturation vapor pressures corresponding to nightside surface temperatures between P$_1$ and P$_3$, that is,
$m_{water} \approx \epsilon [{p_{sat}(P_1) - p_{sat}(P_3)}]/{g}$,
where $\epsilon \approx 0.64$ is the ratio of the molecular weight of water vapor to that of {the background atmosphere}. 

\begin{figure}[!ht]
\centering
\includegraphics[width=0.9\textwidth]{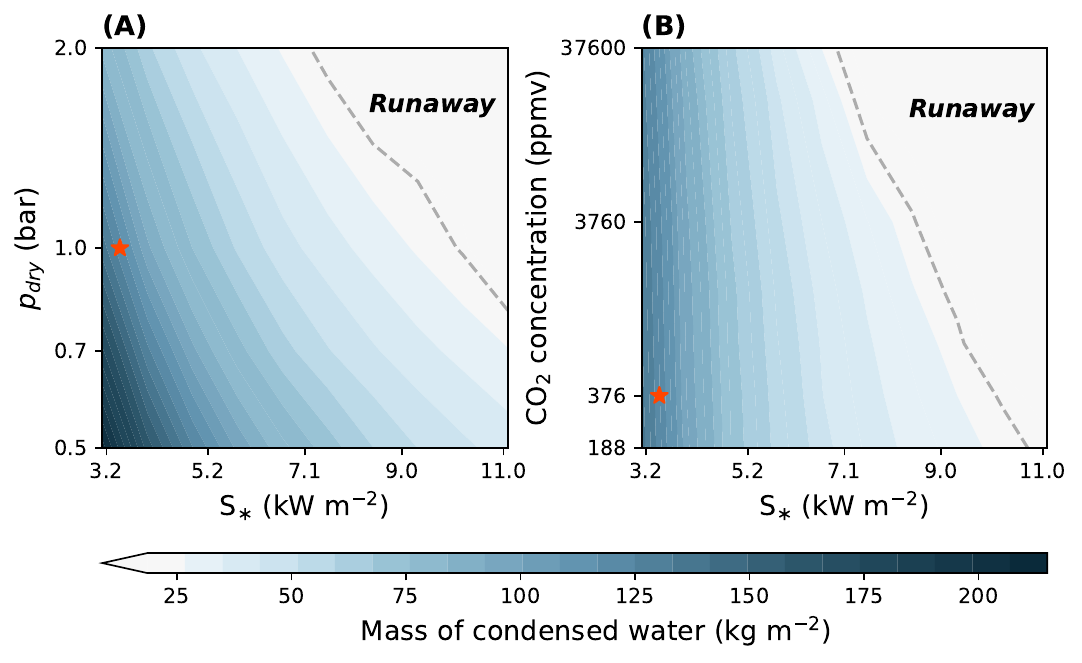}
\caption{Mass of surface water in the collapsed state as a function of  the stellar flux $S_\ast$ and the background surface pressure $p_{dry}$ (A), and a function of $S_\ast$ and the \ce{CO2} concentration $f_{\ce{CO2}}$ (B). In panel(A), $f_{\ce{CO2}} = 376$ ppmv; in panel(B), $p_{dry} = 1$ bar. Gray dashed curves in both panels mark the boundaries of runaway greenhouse state where no surface water can exist, and red stars correspond to the example shown in Figure~\ref{fig:OLR_Tn}. }\label{fig:factors_mwater}
\end{figure}

The divergence between P$_1$ and P$_3$ is significantly influenced by the stellar flux $S_\ast$, the background surface pressure $p_{dry}$, and the  {\ce{CO2} concentration $f_{\ce{CO2}}$}. The latter two can reduce the global mean OLR by enhancing the atmospheric greenhouse effect. 
An increase in any of $S_\ast$, $p_{dry}$, or $f_{\ce{CO2}}$ potentially diminishes this divergence (Figure~\ref{fig:OLR_Tn}A) until P$_1$ and P$_3$ converge at the saddle point $P_2$. Then any further increase in $S_\ast$, $p_{dry}$, or $f_{\ce{CO2}}$ will trigger the Komabayashi-Ingersoll type runaway greenhouse effect, and there can be no water on the night surface. The results in Figure~\ref{fig:factors_mwater} are consistent with this analysis, showing that the mass of surface water in the collapsed state decreases with $S_\ast$, $p_{dry}$, or $f_{\ce{CO2}}$. The equivalent global thickness is less than 20 cm for the parameters surveyed in our MRCS simulations. We also check the phase state of the water on the nightside surface. The nightside surface temperatures \added{in the collapsed state remain below the freezing point of water in our simulations shown in Figure~\ref{fig:factors_mwater}, indicating that the water condenses in the form of ice. }

A global equivalent water depth of only 20 cm is not suitable for the maintenance of long-lived liquid water at the bottom or the edge of ice sheets. \cite{leconte2013} suggested that planets in the Venus zone can acquire more surface water than the values shown in Figure~\ref{fig:factors_mwater} through other processes, such as meteoritic delivery and volcanic outgassing. If water vapor is supplied to the atmosphere slowly enough without triggering the runaway greenhouse effect, the ice sheet on the night surface can grow to about 100 meters thick over long timescales,  {which could support the formation of long-lived liquid water on planets in the Venus zone.}

One issue is that, while water is the predominant outgassed substance in these astrophysical or geological processes, it is not the only species released \citep{gaillard2014theoretical,ortenzi2020mantle,gaillard2021diverse,kite2020exoplanet}. The global mean outgassing flux of \ce{CO2} and \ce{N2} estimated for the present Earth are 0.089 and 9.5$\times 10^{-4}$ relative to that of \ce{H2O}, respectively \citep[see][Chapter 7.2.2]{catling2017evulotionbook}. Using this Earth-like outgassing composition, 4 bars of \ce{CO2} and 0.03 bar of \ce{N2} would be released into the atmosphere along with water vapor when the nightside ice sheet grows to 100 meters thick. The gray dashed curves in Figure~\ref{fig:factors_mwater} indicate that the increase of background surface pressure or  {\ce{CO2} concentration to critical levels} can trigger the Komabayashi-Ingersoll type runaway greenhouse effect on Venus zone planets.  {Assuming the limiting case that all outgassed \ce{CO2} stays in the atmosphere,} 4 bar \ce{CO2} would be very likely to force the planet to the runaway greenhouse state  {by extrapolating} our MRCS simulations in Figure~\ref{fig:factors_mwater}. As a result, we propose a critical condition for the formation of thick water ice layers on the night surface, namely that species released into the atmosphere, in addition to water vapor, must also undergo condensation on the night surface. This process requires the presence of a very tenuous gaseous envelope around the planet \citep{Wordsworth2015,koll2016,ouyang2024}. 

\section{Discussion} \label{sec:discussion}
We develop a MRCS model to study the climate transition of Venus zone planets in a two column framework. We show that the retention of surface water on the night surface is associated with the climate transition from an unstable  {equilibrium} state to a stable state under the same  {stellar flux}. Our simulations indicate that  {the surface water mass declines with the incident stellar flux, background surface pressure, and optical thickness of non-condensible greenhouse gases.} Only planets in the Venus zone with very tenuous atmospheres are likely to form substantial ice sheets on their night surfaces.

Our approach provides a straightforward way to evaluate climate conditions on exoplanets in the Venus zone around M dwarfs under various conditions. Today, characterizing the surface and climate conditions of these planets is technically feasible via transmission and secondary eclipse spectroscopy \citep{Morley_2017temperate,wordsworth2022atmospheres,Way_2023synergies}. Several exoplanets in the Venus zone around M dwarfs have been proposed as important targets for atmospheric characterization \citep{Ostberg_2023venuszone}. Among those, preliminary observations by the James Webb Space Telescope indicate that TRAPPIST-1 b and c, GJ 1132b lack a thick gaseous envelope \citep{Greene_2023trappist1b, Zieba_2023trappist1c, Xue_2024gj1132b}. In contrast, L 98-59 d may possess a secondary atmosphere with possible sulfur species \citep{Banerjee_2024l98-59d,Gressier_2024l98-59d}, which might suggest active volcanic activity on the planet. Recently, the Transiting Exoplanet Survey Satellite (TESS) has also discovered one temperate Earth-sized planet LP 791-18d orbiting an M6 star with possible strong volcanic activity at the surface \citep{peterson2023temperate}. 
Although volcanic outgassing plays an essential role in preventing terrestrial planets in the habitable zone from being trapped in the snowball state \citep{hoffman1998snowball,pierrehumbert2004snowballco2}, our study indicates that it can easily  {force planets in the Venus zone to enter the runaway greenhouse state}.

Lastly, we address several factors that were ignored by our MRCS model but should not alter the general conclusion in this paper. One factor is the assumption of our two-column framework that the atmospheric circulation on tidally locked planets in the Venus zone is dominated by the thermally direct overturning circulation between the day and night columns. The Venus zone planets around M dwarfs usually rotate fast enough that their Rossby deformation radii are comparable to their planetary radii, so a relatively large temperature difference can develop in the night hemisphere, especially when the atmosphere is hot and thin \citep{leconte2013,koll2016}. The night column in our MRCS model can only reflect the night hemispheric mean state, but is slightly warmer than the actual permanent surface cold trap. 

Another important factor ignored by our MRCS model is the cloud radiative effect. In our MRCS simulations, the tropopause is subsaturated (see relative humidity profiles in Figure~\ref{fig:OLR_Tn}C and E), which is consistent with our assumption that water vapor is only cold-trapped by the surface and is well mixed in the atmosphere. However, our MRCS model assumes that the atmosphere is gray in the thermal infrared  {and transparent in the shortwave}. The nongray radiative property of real gases in the thermal infrared can efficiently cool the upper atmosphere \citep{leconte2013runawaynature} and trigger high-level cloud formation near the tropopause, while the atmospheric absorption of incoming stellar flux in the shortwave spectrum can lead to an opposing effect. The cloud warming effect has been proposed to play a significant role in preventing ocean formation from a thick steam atmosphere on early Venus \citep{turbet2021venuscloud}. For tidally locked planets in the Venus zone around M dwarfs, possibility of nightside cloud formation in a much thinner atmosphere and its role in surface water condensation cannot be answered with our two-column model and should be explored by complex GCMs in future. 


\begin{acknowledgments}
 {The authors thank the referee for thoughtful comments that improved the manuscript.} The authors thank Daniel Koll for sharing his code of the radiative-convective-subsiding model and Chengyang Luo for helpful discussions. F.D. acknowledge funding support from the Fundamental Research Funds for the Central Universities (Peking University).  {Our moist radiative-convective-subsiding model and outputs are available on Github at \url{https://github.com/OuyangYueyun/collapse_runaway}}.
\end{acknowledgments}

\begin{contribution}
F.D. conceived the original idea and supervised this project. Y.O. developed the moist radiative-convective-subsiding model and performed the numerical simulations. Y.O. and F.D. analyzed the simulation results and wrote the paper. All authors discussed the results and implications.


\end{contribution}

%



\appendix

\section{Infrared opacity of greenhouse gases} \label{sec:calc_tau}

In our MRCS model, the water vapor is well mixed in the atmosphere and is in phase equilibrium with the nightside surface water. The saturation vapor pressure of water is calculated by the Clausius-Clapeyron relation as
\begin{equation}
    p_{sat,\ce{H2O}}=\frac{\exp(43.494-\frac{6545.8}{t+278})}{(t+868)^2}\qquad(t\leq0^\circ \rm{C})
\end{equation}
for temperatures below $0^\circ$C, and
\begin{equation}
    p_{sat,\ce{H2O}}=\frac{\exp(34.494-\frac{4924.99}{t+237.1})}{(t+105)^{1.57}}\qquad(t>0^\circ \rm{C})
\end{equation}
for temperatures above $0^\circ$C, where $p_{sat,\ce{H2O}}$ is the saturation vapor pressure of water in Pa and $t$ is the surface water temperature in $0^\circ$C \citep{huang2018simple}. Then the saturation mass fraction of water vapor at the nightside surface can be calculated as \begin{equation}
q_{sat} = \frac{\epsilon p_{sat,\ce{H2O}} / p_{dry}}{1+\epsilon p_{sat,\ce{H2O}} / p_{dry}},
\end{equation}
where $\epsilon$ is the ratio of molecular weight of water vapor to that of the background atmosphere, and $p_{dry}$ is the surface partial pressure of the background atmosphere. When water vapor is dilute in the atmosphere, $q_{sat} \approx \epsilon p_{sat,\ce{H2O}} / p_{s}$. 

Eq.~\ref{eq:tauwv} shows how the total optical thickness of water vapor in thermal infrared $\tau_{wv}$ is related to $q_{sat}$ and the mean absorption cross section $\bar{\kappa}_{wv}$. First, we use the line-by-line radiation code PyRADS to calculate the spectral distribution of the absorption coefficient $\kappa_{\nu}$. The code relies on the HITRAN 2016 molecular spectroscopic database \citep{gordon2017hitran2016}, and also uses the Mlawer-Tobin-Clough-Kneizys-Davies (MTCKD) model to compute self and foreign continuum absorption of water vapor \citep{mlawer2012}. The spectral range in our calculation is from 1 to 3500 cm$^{-1}$, with a resolution of 0.01 cm$^{-1}$. Next, we calculate the Rosseland mean value to represent the gray absorption coefficient of water vapor in the MRCS model, with the formula
\begin{equation}
    \frac{1}{\overline{\kappa}_{wv}}=\frac{\int^{\infty}_0\frac{1}{\kappa_\nu}\frac{dB_\nu(T)}{dT}d\nu}{\int^{\infty}_0\frac{dB_\nu(T)}{dT}d\nu},
\end{equation}
where $\nu$ is the frequency, and $B_\nu$ is the Planck function. For our simulations in which water vapor begins to condense on the night surface, the night surface temperature is $\sim 300$ K. The thermal infrared absorption feature of water vapor with such a high abundance is optically opaque in most spectral regions with weakly absorption window regions near 10 $\mu$m, where the Rosseland mean is applicable \citep{thomas2002rt}. 

We also include the greenhouse effect of \ce{CO2} in our MRCS model. The thermal infrared absorption feature of \ce{CO2} with the abundance in our simulations is very different from that of water vapor, which is transparent in most spectral regions with strong absorption bands near 15 $\mu$m, where the Planck mean is applicable \citep{thomas2002rt}. We calculate the mean optical thickness of \ce{CO2} $\bar{\tau}_{\ce{CO2}}$ following Eq.~4 in \citet{koll2019secondary}
\begin{equation}
    \bar{\tau}_{\ce{CO2}} = - \ln \left( \frac{\int^{\infty}_0 \exp(-\tau_{\nu,\ce{CO2}}) {B_\nu(T)} d\nu}{\int^{\infty}_0  {B_\nu(T)} d\nu} \right).
\end{equation}

\section{Comparison with previous 3D GCM results}\label{sec:reproduce}

\cite{leconte2013} used an upgraded version of the LMD generic Global Climate Model with 3D dynamic core and correlated-k method to find the critical global mean value of water vapor ($m_v$) that triggers surface water condensation on close-in synchronously rotating planets. Their simulations assume an Earth-like atmosphere with \ce{N2} and 376 ppmv \ce{CO2}. The surface partial pressure of \ce{N2} is varied, and the corresponding $m_v$ is given in Figure~10 in \cite{leconte2013}. 

\begin{figure}[!ht]
\centering
\includegraphics[width=0.8\textwidth]{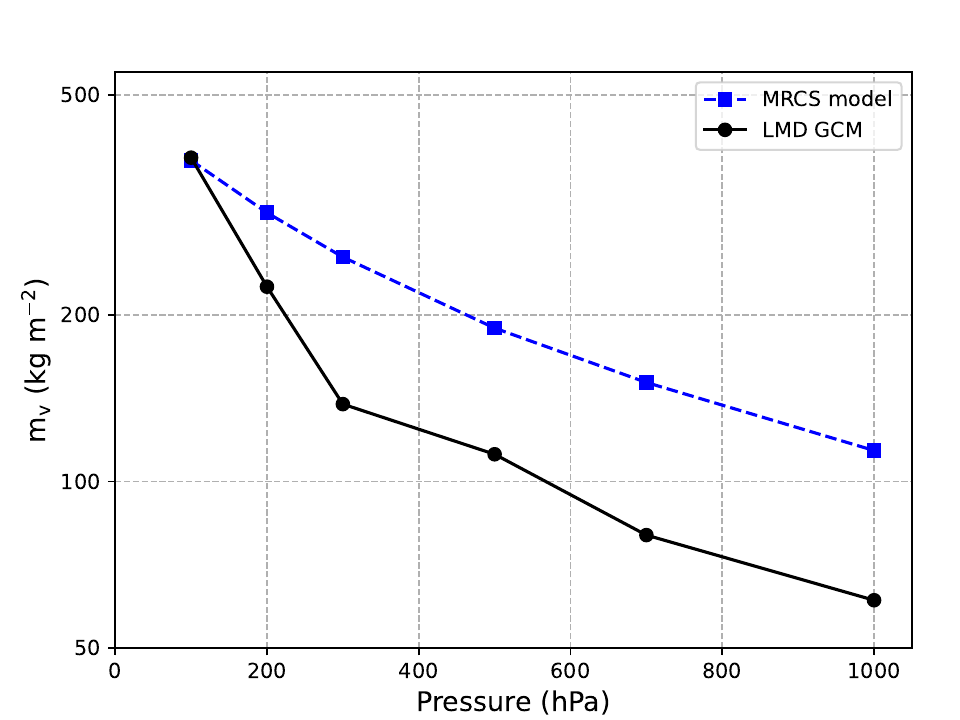}
\caption{Critical global mean value of water vapor ($m_v$) that triggers surface water condensation on the close-in synchronously rotating planet Gl~581c, when the surface partial pressure of \ce{N2} is varied from 100 hPa to 1000 hPa. Black solid curve with solid circles is estimated by GCM simulations in \cite{leconte2013}, and blue dashed curve with squares is estimated by our MRCS model. }\label{fig:leconte2013}
\end{figure}

The unstable equilibrium state found in our MRCS model \added{(e.g., \ce{P1} in Figure~\ref{fig:OLR_Tn}A)} is directly related to the critical value of $m_v$ found in \cite{leconte2013}. For example, the column-integrated amount of water vapor \added{in the unstable equilibrium state \ce{P1} in Figure~\ref{fig:OLR_Tn}C} gives the critical value of $m_v$ when $p_{dry}$ = 1 bar and $f_{\ce{CO2}}$ = 376 ppmv. The critical values of $m_v$ found in our MRCS model are consistent with those in \cite{leconte2013}, as shown in Figure~\ref{fig:leconte2013}. The small deviations might be related to the cloud radiative effect, which is discussed in Section~\ref{sec:discussion}.

\bibliography{sample7}{}

\begin{thebibliography}{}
\expandafter\ifx\csname natexlab\endcsname\relax\def\natexlab#1{#1}\fi
\providecommand{\url}[1]{\href{#1}{#1}}
\providecommand{\dodoi}[1]{doi:~\href{http://doi.org/#1}{\nolinkurl{#1}}}
\providecommand{\doeprint}[1]{\href{http://ascl.net/#1}{\nolinkurl{http://ascl.net/#1}}}
\providecommand{\doarXiv}[1]{\href{https://arxiv.org/abs/#1}{\nolinkurl{https://arxiv.org/abs/#1}}}

\bibitem[{Y. {Abe} {et~al.}(2011){Abe}, {Abe-Ouchi}, {Sleep}, \& {Zahnle}}]{abe2011habitable}
{Abe}, Y., {Abe-Ouchi}, A., {Sleep}, N.~H., \& {Zahnle}, K.~J. 2011, \bibinfo{title}{{Habitable Zone Limits for Dry Planets},} Astrobiology, 11, 443, \dodoi{10.1089/ast.2010.0545}

\bibitem[{P. Auclair-Desrotour \& K. Heng(2020)Auclair-Desrotour \& Heng}]{Auclair_Desrotour_2020twocolomn}
Auclair-Desrotour, P., \& Heng, K. 2020, \bibinfo{title}{Atmospheric stability and collapse on tidally locked rocky planets,} Astronomy \& Astrophysics, 638, A77, \dodoi{10.1051/0004-6361/202037513}

\bibitem[{A. Banerjee {et~al.}(2024)Banerjee, Barstow, Gressier, Espinoza, Sing, Allen, Birkmann, Challener, Crouzet, Haswell, Lewis, Lewis, \& Yang}]{Banerjee_2024l98-59d}
Banerjee, A., Barstow, J.~K., Gressier, A., {et~al.} 2024, \bibinfo{title}{Atmospheric Retrievals Suggest the Presence of a Secondary Atmosphere and Possible Sulfur Species on L98-59 d from JWST Nirspec G395H Transmission Spectroscopy,} The Astrophysical Journal Letters, 975, L11, \dodoi{10.3847/2041-8213/ad73d0}

\bibitem[{R. {Barnes}(2017){Barnes}}]{barnes2017tidal}
{Barnes}, R. 2017, \bibinfo{title}{{Tidal locking of habitable exoplanets},} Celestial Mechanics and Dynamical Astronomy, 129, 509, \dodoi{10.1007/s10569-017-9783-7}

\bibitem[{D.~C. {Catling} \& J.~F. {Kasting}(2017){Catling} \& {Kasting}}]{catling2017evulotionbook}
{Catling}, D.~C., \& {Kasting}, J.~F. 2017, {Atmospheric Evolution on Inhabited and Lifeless Worlds} (Cambridge University Press)

\bibitem[{F. {Ding} \& R.~D. {Wordsworth}(2020){Ding} \& {Wordsworth}}]{ding2020stabilization}
{Ding}, F., \& {Wordsworth}, R.~D. 2020, \bibinfo{title}{{Stabilization of Dayside Surface Liquid Water via Tropopause Cold Trapping on Arid Slowly Rotating Tidally Locked Planets},} \apjl, 891, L18, \dodoi{10.3847/2041-8213/ab77d1}

\bibitem[{F. Ding \& R.~D. Wordsworth(2021)Ding \& Wordsworth}]{Ding_2021lastsaturation}
Ding, F., \& Wordsworth, R.~D. 2021, \bibinfo{title}{Multiple Moist Climate Equilibrium States on Arid Rocky M-dwarf Planets: A Last-saturation Tracer Analysis,} The Planetary Science Journal, 2, 201, \dodoi{10.3847/psj/ac2236}

\bibitem[{F. {Ding} \& R.~D. {Wordsworth}(2022){Ding} \& {Wordsworth}}]{ding2022prospects}
{Ding}, F., \& {Wordsworth}, R.~D. 2022, \bibinfo{title}{{Prospects for Water Vapor Detection in the Atmospheres of Temperate and Arid Rocky Exoplanets around M-dwarf Stars},} \apjl, 925, L8, \dodoi{10.3847/2041-8213/ac4a5d}

\bibitem[{F. {Gaillard} \& B. {Scaillet}(2014){Gaillard} \& {Scaillet}}]{gaillard2014theoretical}
{Gaillard}, F., \& {Scaillet}, B. 2014, \bibinfo{title}{{A theoretical framework for volcanic degassing chemistry in a comparative planetology perspective and implications for planetary atmospheres},} Earth and Planetary Science Letters, 403, 307, \dodoi{10.1016/j.epsl.2014.07.009}

\bibitem[{F. {Gaillard} {et~al.}(2021){Gaillard}, {Bouhifd}, {F{\"u}ri}, {Malavergne}, {Marrocchi}, {Noack}, {Ortenzi}, {Roskosz}, \& {Vulpius}}]{gaillard2021diverse}
{Gaillard}, F., {Bouhifd}, M.~A., {F{\"u}ri}, E., {et~al.} 2021, \bibinfo{title}{{The Diverse Planetary Ingassing/Outgassing Paths Produced over Billions of Years of Magmatic Activity},} \ssr, 217, 22, \dodoi{10.1007/s11214-021-00802-1}

\bibitem[{C. {Goldblatt} \& A.~J. {Watson}(2012){Goldblatt} \& {Watson}}]{goldblatt2012runaway}
{Goldblatt}, C., \& {Watson}, A.~J. 2012, \bibinfo{title}{{The runaway greenhouse: implications for future climate change, geoengineering and planetary atmospheres},} Philosophical Transactions of the Royal Society of London Series A, 370, 4197, \dodoi{10.1098/rsta.2012.0004}

\bibitem[{P. {Goldreich} \& S. {Peale}(1966){Goldreich} \& {Peale}}]{goldreich1966spinorbit}
{Goldreich}, P., \& {Peale}, S. 1966, \bibinfo{title}{{Spin-orbit coupling in the solar system},} Astronomical Journal, 71, 425, \dodoi{10.1086/109947}

\bibitem[{I.~E. {Gordon} {et~al.}(2017){Gordon}, {Rothman}, {Hill}, {Kochanov}, {Tan}, {Bernath}, {Birk}, {Boudon}, {Campargue}, {Chance}, {Drouin}, {Flaud}, {Gamache}, {Hodges}, {Jacquemart}, {Perevalov}, {Perrin}, {Shine}, {Smith}, {Tennyson}, {Toon}, {Tran}, {Tyuterev}, {Barbe}, {Cs{\'a}sz{\'a}r}, {Devi}, {Furtenbacher}, {Harrison}, {Hartmann}, {Jolly}, {Johnson}, {Karman}, {Kleiner}, {Kyuberis}, {Loos}, {Lyulin}, {Massie}, {Mikhailenko}, {Moazzen-Ahmadi}, {M{\"u}ller}, {Naumenko}, {Nikitin}, {Polyansky}, {Rey}, {Rotger}, {Sharpe}, {Sung}, {Starikova}, {Tashkun}, {Auwera}, {Wagner}, {Wilzewski}, {Wcis{\l}o}, {Yu}, \& {Zak}}]{gordon2017hitran2016}
{Gordon}, I.~E., {Rothman}, L.~S., {Hill}, C., {et~al.} 2017, \bibinfo{title}{{The HITRAN2016 molecular spectroscopic database},} \jqsrt, 203, 3, \dodoi{10.1016/j.jqsrt.2017.06.038}

\bibitem[{T.~P. Greene {et~al.}(2023)Greene, Bell, Ducrot, Dyrek, Lagage, \& Fortney}]{Greene_2023trappist1b}
Greene, T.~P., Bell, T.~J., Ducrot, E., {et~al.} 2023, \bibinfo{title}{Thermal emission from the Earth-sized exoplanet TRAPPIST-1 b using JWST,} Nature, 618, 39–42, \dodoi{10.1038/s41586-023-05951-7}

\bibitem[{A. Gressier {et~al.}(2024)Gressier, Espinoza, Allen, Sing, Banerjee, Barstow, Valenti, Lewis, Birkmann, Challener, Manjavacas, Alves~de Oliveira, Crouzet, \& Beck}]{Gressier_2024l98-59d}
Gressier, A., Espinoza, N., Allen, N.~H., {et~al.} 2024, \bibinfo{title}{Hints of a Sulfur-rich Atmosphere around the 1.6 R$_\oplus$ Super-Earth L98-59 d from JWST NIRspec G395H Transmission Spectroscopy,} The Astrophysical Journal Letters, 975, L10, \dodoi{10.3847/2041-8213/ad73d1}

\bibitem[{R. {Heller} {et~al.}(2011){Heller}, {Leconte}, \& {Barnes}}]{heller2011tidal}
{Heller}, R., {Leconte}, J., \& {Barnes}, R. 2011, \bibinfo{title}{{Tidal obliquity evolution of potentially habitable planets},} \aap, 528, A27, \dodoi{10.1051/0004-6361/201015809}

\bibitem[{P.~F. {Hoffman} {et~al.}(1998){Hoffman}, {Kaufman}, {Halverson}, \& {Schrag}}]{hoffman1998snowball}
{Hoffman}, P.~F., {Kaufman}, A.~J., {Halverson}, G.~P., \& {Schrag}, D.~P. 1998, \bibinfo{title}{{A Neoproterozoic Snowball Earth},} Science, 281, 1342, \dodoi{10.1126/science.281.5381.1342}

\bibitem[{J. Huang(2018)Huang}]{huang2018simple}
Huang, J. 2018, \bibinfo{title}{A Simple Accurate Formula for Calculating Saturation Vapor Pressure of Water and Ice,} Journal of Applied Meteorology and Climatology, 57, 1265 , \dodoi{10.1175/JAMC-D-17-0334.1}

\bibitem[{A.~P. {Ingersoll}(1969){Ingersoll}}]{ingersoll1969runaway}
{Ingersoll}, A.~P. 1969, \bibinfo{title}{{The Runaway Greenhouse: A History of Water on Venus.},} Journal of the Atmospheric Sciences, 26, 1191, \dodoi{10.1175/1520-0469(1969)026<1191:TRGAHO>2.0.CO;2}

\bibitem[{S.~R. Kane {et~al.}(2014)Kane, Kopparapu, \& Domagal-Goldman}]{Kane_2014venuszone}
Kane, S.~R., Kopparapu, R.~K., \& Domagal-Goldman, S.~D. 2014, \bibinfo{title}{ON THE FREQUENCY OF POTENTIAL VENUS ANALOGS FROM KEPLER DATA,} The Astrophysical Journal, 794, L5, \dodoi{10.1088/2041-8205/794/1/l5}

\bibitem[{S.~R. Kane \& K. von Braun(2008)Kane \& von Braun}]{Kane_2008transit}
Kane, S.~R., \& von Braun, K. 2008, \bibinfo{title}{Constraining Orbital Parameters through Planetary Transit Monitoring,} The Astrophysical Journal, 689, 492–498, \dodoi{10.1086/592381}

\bibitem[{J.~F. {Kasting} {et~al.}(1993){Kasting}, {Whitmire}, \& {Reynolds}}]{kasting1993habitable}
{Kasting}, J.~F., {Whitmire}, D.~P., \& {Reynolds}, R.~T. 1993, \bibinfo{title}{{Habitable Zones around Main Sequence Stars},} \icarus, 101, 108, \dodoi{10.1006/icar.1993.1010}

\bibitem[{E.~S. {Kite} \& M.~N. {Barnett}(2020){Kite} \& {Barnett}}]{kite2020exoplanet}
{Kite}, E.~S., \& {Barnett}, M.~N. 2020, \bibinfo{title}{{Exoplanet secondary atmosphere loss and revival},} Proceedings of the National Academy of Science, 117, 18264, \dodoi{10.1073/pnas.2006177117}

\bibitem[{T. Kodama {et~al.}(2018)Kodama, Nitta, Genda, Takao, O’ishi, Abe‐Ouchi, \& Abe}]{Kodama_2018dryplanet}
Kodama, T., Nitta, A., Genda, H., {et~al.} 2018, \bibinfo{title}{Dependence of the Onset of the Runaway Greenhouse Effect on the Latitudinal Surface Water Distribution of Earth‐Like Planets,} Journal of Geophysical Research: Planets, 123, 559–574, \dodoi{10.1002/2017je005383}

\bibitem[{D.~D. Koll \& D.~S. Abbot(2016)Koll \& Abbot}]{koll2016}
Koll, D.~D., \& Abbot, D.~S. 2016, \bibinfo{title}{Temperature structure and atmospheric circulation of dry tidally locked rocky exoplanets,} The Astrophysical Journal, 825, 99, \dodoi{10.3847/0004-637X/825/2/99}

\bibitem[{D.~D.~B. Koll \& T.~W. Cronin(2018)Koll \& Cronin}]{koll2018}
Koll, D. D.~B., \& Cronin, T.~W. 2018, \bibinfo{title}{Earth\&\#x2019;s outgoing longwave radiation linear due to H<sub>2</sub>O greenhouse effect,} Proceedings of the National Academy of Sciences, 115, 10293, \dodoi{10.1073/pnas.1809868115}

\bibitem[{D.~D.~B. Koll \& T.~W. Cronin(2019)Koll \& Cronin}]{Koll_2019}
Koll, D. D.~B., \& Cronin, T.~W. 2019, \bibinfo{title}{Hot Hydrogen Climates Near the Inner Edge of the Habitable Zone,} The Astrophysical Journal, 881, 120, \dodoi{10.3847/1538-4357/ab30c4}

\bibitem[{D.~D.~B. {Koll} {et~al.}(2019){Koll}, {Malik}, {Mansfield}, {Kempton}, {Kite}, {Abbot}, \& {Bean}}]{koll2019secondary}
{Koll}, D. D.~B., {Malik}, M., {Mansfield}, M., {et~al.} 2019, \bibinfo{title}{{Identifying Candidate Atmospheres on Rocky M Dwarf Planets via Eclipse Photometry},} \apj, 886, 140, \dodoi{10.3847/1538-4357/ab4c91}

\bibitem[{M. {Komabayasi}(1967){Komabayasi}}]{komabayashi1967runaway}
{Komabayasi}, M. 1967, \bibinfo{title}{{Discrete Equilibrium Temperatures of a Hypothetical Planet with the Atmosphere and the Hydrosphere of One Component-Two Phase System under Constant Solar Radiation},} Journal of the Meteorological Society of Japan, 45, 137, \dodoi{10.2151/jmsj1965.45.1_137}

\bibitem[{J. {Leconte} {et~al.}(2013{\natexlab{a}}){Leconte}, {Forget}, {Charnay}, {Wordsworth}, \& {Pottier}}]{leconte2013runawaynature}
{Leconte}, J., {Forget}, F., {Charnay}, B., {Wordsworth}, R., \& {Pottier}, A. 2013{\natexlab{a}}, \bibinfo{title}{{Increased insolation threshold for runaway greenhouse processes on Earth-like planets},} \nat, 504, 268, \dodoi{10.1038/nature12827}

\bibitem[{J. {Leconte} {et~al.}(2013{\natexlab{b}}){Leconte}, {Forget}, {Charnay}, {Wordsworth}, {Selsis}, {Millour}, \& {Spiga}}]{leconte2013}
{Leconte}, J., {Forget}, F., {Charnay}, B., {et~al.} 2013{\natexlab{b}}, \bibinfo{title}{{3D climate modeling of close-in land planets: Circulation patterns, climate moist bistability, and habitability},} \aap, 554, A69, \dodoi{10.1051/0004-6361/201321042}

\bibitem[{A.~H. {Lobo} {et~al.}(2023){Lobo}, {Shields}, {Palubski}, \& {Wolf}}]{lobo2023terminator}
{Lobo}, A.~H., {Shields}, A.~L., {Palubski}, I.~Z., \& {Wolf}, E. 2023, \bibinfo{title}{{Terminator Habitability: The Case for Limited Water Availability on M-dwarf Planets},} \apj, 945, 161, \dodoi{10.3847/1538-4357/aca970}

\bibitem[{E.~J. Mlawer {et~al.}(2012)Mlawer, Payne, Moncet, Delamere, Alvarado, \& Tobin}]{mlawer2012}
Mlawer, E.~J., Payne, V.~H., Moncet, J.-L., {et~al.} 2012, \bibinfo{title}{Development and recent evaluation of the MT\_CKD model of continuum absorption,} Philosophical Transactions of the Royal Society A: Mathematical, Physical and Engineering Sciences, 370, 2520, \dodoi{https://doi.org/10.1098/rsta.2011.0295}

\bibitem[{C.~V. Morley {et~al.}(2017)Morley, Kreidberg, Rustamkulov, Robinson, \& Fortney}]{Morley_2017temperate}
Morley, C.~V., Kreidberg, L., Rustamkulov, Z., Robinson, T., \& Fortney, J.~J. 2017, \bibinfo{title}{Observing the Atmospheres of Known Temperate Earth-sized Planets with JWST,} The Astrophysical Journal, 850, 121, \dodoi{10.3847/1538-4357/aa927b}

\bibitem[{S. {Nakajima} {et~al.}(1992){Nakajima}, {Hayashi}, \& {Abe}}]{nakajima1992runaway}
{Nakajima}, S., {Hayashi}, Y.-Y., \& {Abe}, Y. 1992, \bibinfo{title}{{A study on the 'runaway greenhouse effect' with a one-dimensional radiative-convective equilibrium model},} Journal of the Atmospheric Sciences, 49, 2256, \dodoi{10.1175/1520-0469(1992)049<2256:ASOTGE>2.0.CO;2}

\bibitem[{L. {Ojha} {et~al.}(2022){Ojha}, {Troncone}, {Buffo}, {Journaux}, \& {McDonald}}]{ojha2022liquid}
{Ojha}, L., {Troncone}, B., {Buffo}, J., {Journaux}, B., \& {McDonald}, G. 2022, \bibinfo{title}{{Liquid water on cold exo-Earths via basal melting of ice sheets},} Nature Communications, 13, 7521, \dodoi{10.1038/s41467-022-35187-4}

\bibitem[{G. {Ortenzi} {et~al.}(2020){Ortenzi}, {Noack}, {Sohl}, {Guimond}, {Grenfell}, {Dorn}, {Schmidt}, {Vulpius}, {Katyal}, {Kitzmann}, \& {Rauer}}]{ortenzi2020mantle}
{Ortenzi}, G., {Noack}, L., {Sohl}, F., {et~al.} 2020, \bibinfo{title}{{Mantle redox state drives outgassing chemistry and atmospheric composition of rocky planets},} Scientific Reports, 10, 10907, \dodoi{10.1038/s41598-020-67751-7}

\bibitem[{C. Ostberg {et~al.}(2023)Ostberg, Kane, Li, Schwieterman, Hill, Bott, Dalba, Fetherolf, Head, \& Unterborn}]{Ostberg_2023venuszone}
Ostberg, C., Kane, S.~R., Li, Z., {et~al.} 2023, \bibinfo{title}{The Demographics of Terrestrial Planets in the Venus Zone,} The Astronomical Journal, 165, 168, \dodoi{10.3847/1538-3881/acbfaf}

\bibitem[{Y. Ouyang \& F. Ding(2024)Ouyang \& Ding}]{ouyang2024}
Ouyang, Y., \& Ding, F. 2024, \bibinfo{title}{Potential surface ice distribution on close-in terrestrial exoplanets around M dwarfs,} Monthly Notices of the Royal Astronomical Society, 531, 251, \dodoi{10.1093/mnras/stae1200}

\bibitem[{M.~S. {Peterson} {et~al.}(2023){Peterson}, {Benneke}, {Collins}, {Piaulet}, {Crossfield}, {Ali-Dib}, {Christiansen}, {Gagn{\'e}}, {Faherty}, {Kite}, {Dressing}, {Charbonneau}, {Murgas}, {Cointepas}, {Almenara}, {Bonfils}, {Kane}, {Werner}, {Gorjian}, {Roy}, {Shporer}, {Pozuelos}, {Socia}, {Cloutier}, {Dietrich}, {Irwin}, {Weiss}, {Waalkes}, {Berta-Thomson}, {Evans}, {Apai}, {Parviainen}, {Pall{\'e}}, {Narita}, {Howard}, {Dragomir}, {Barkaoui}, {Gillon}, {Jehin}, {Ducrot}, {Benkhaldoun}, {Fukui}, {Mori}, {Nishiumi}, {Kawauchi}, {Ricker}, {Latham}, {Winn}, {Seager}, {Isaacson}, {Bixel}, {Gibbs}, {Jenkins}, {Smith}, {Chavez}, {Rackham}, {Henning}, {Gabor}, {Chen}, {Espinoza}, {Jensen}, {Collins}, {Schwarz}, {Conti}, {Wang}, {Kielkopf}, {Mao}, {Horne}, {Sefako}, {Quinn}, {Moldovan}, {Fausnaugh}, {F{\.z}{\.z}r{\'e}sz}, \& {Barclay}}]{peterson2023temperate}
{Peterson}, M.~S., {Benneke}, B., {Collins}, K., {et~al.} 2023, \bibinfo{title}{{A temperate Earth-sized planet with tidal heating transiting an M6 star},} \nat, 617, 701, \dodoi{10.1038/s41586-023-05934-8}

\bibitem[{R.~T. {Pierrehumbert}(2004){Pierrehumbert}}]{pierrehumbert2004snowballco2}
{Pierrehumbert}, R.~T. 2004, \bibinfo{title}{{High levels of atmospheric carbon dioxide necessary for the termination of global glaciation},} \nat, 429, 646, \dodoi{10.1038/nature02640}

\bibitem[{R.~T. {Pierrehumbert}(2010){Pierrehumbert}}]{pierrehumbert2010ppcbook}
{Pierrehumbert}, R.~T. 2010, {Principles of Planetary Climate} (Cambridge University Press)

\bibitem[{G.~C. {Simpson}(1928){Simpson}}]{simpson1928runaway}
{Simpson}, G.~C. 1928, \bibinfo{title}{{Further Studies in Terrestrial RADIATION1},} Monthly Weather Review, 56, 322, \dodoi{10.1175/1520-0493(1928)56<322:FSITR>2.0.CO;2}

\bibitem[{G.~E. {Thomas} \& K. {Stamnes}(2002){Thomas} \& {Stamnes}}]{thomas2002rt}
{Thomas}, G.~E., \& {Stamnes}, K. 2002, {Radiative Transfer in the Atmosphere and Ocean}

\bibitem[{M. {Turbet} {et~al.}(2021){Turbet}, {Bolmont}, {Chaverot}, {Ehrenreich}, {Leconte}, \& {Marcq}}]{turbet2021venuscloud}
{Turbet}, M., {Bolmont}, E., {Chaverot}, G., {et~al.} 2021, \bibinfo{title}{{Day-night cloud asymmetry prevents early oceans on Venus but not on Earth},} \nat, 598, 276, \dodoi{10.1038/s41586-021-03873-w}

\bibitem[{A. {Wandel}(2023){Wandel}}]{wandel2023habitability}
{Wandel}, A. 2023, \bibinfo{title}{{Habitability and sub glacial liquid water on planets of M-dwarf stars},} Nature Communications, 14, 2125, \dodoi{10.1038/s41467-023-37487-9}

\bibitem[{M.~J. Way {et~al.}(2023)Way, Ostberg, Foley, Gillmann, Höning, Lammer, O’Rourke, Persson, Plesa, Salvador, Scherf, \& Weller}]{Way_2023synergies}
Way, M.~J., Ostberg, C., Foley, B.~J., {et~al.} 2023, \bibinfo{title}{Synergies Between Venus \& Exoplanetary Observations: Venus and Its Extrasolar Siblings,} Space Science Reviews, 219, \dodoi{10.1007/s11214-023-00953-3}

\bibitem[{J.~N. {Winn}(2010){Winn}}]{winn2010exop.book.transit}
{Winn}, J.~N. 2010, in Exoplanets, ed. S.~{Seager} (The University of Arizona Press), 55--77, \dodoi{10.48550/arXiv.1001.2010}

\bibitem[{R. Wordsworth(2015)Wordsworth}]{Wordsworth2015}
Wordsworth, R. 2015, \bibinfo{title}{Atmospheric heat redistribution and collapse on tidally locked rocky planets,} The Astrophysical Journal, 806, 180, \dodoi{10.1088/0004-637X/806/2/180}

\bibitem[{R. {Wordsworth} \& L. {Kreidberg}(2022){Wordsworth} \& {Kreidberg}}]{wordsworth2022atmospheres}
{Wordsworth}, R., \& {Kreidberg}, L. 2022, \bibinfo{title}{{Atmospheres of Rocky Exoplanets},} \araa, 60, 159, \dodoi{10.1146/annurev-astro-052920-125632}

\bibitem[{Q. Xue {et~al.}(2024)Xue, Bean, Zhang, Mahajan, Ih, Eastman, Lunine, Mansfield, Coy, Kempton, Koll, \& Kite}]{Xue_2024gj1132b}
Xue, Q., Bean, J.~L., Zhang, M., {et~al.} 2024, \bibinfo{title}{JWST Thermal Emission of the Terrestrial Exoplanet GJ 1132b,} The Astrophysical Journal Letters, 973, L8, \dodoi{10.3847/2041-8213/ad72e9}

\bibitem[{J. {Yang} {et~al.}(2019){Yang}, {Abbot}, {Koll}, {Hu}, \& {Showman}}]{yang2019oceaninneredge}
{Yang}, J., {Abbot}, D.~S., {Koll}, D. D.~B., {Hu}, Y., \& {Showman}, A.~P. 2019, \bibinfo{title}{{Ocean Dynamics and the Inner Edge of the Habitable Zone for Tidally Locked Terrestrial Planets},} \apj, 871, 29, \dodoi{10.3847/1538-4357/aaf1a8}

\bibitem[{J. Yang {et~al.}(2013)Yang, Cowan, \& Abbot}]{Yang_2013cloud}
Yang, J., Cowan, N.~B., \& Abbot, D.~S. 2013, \bibinfo{title}{STABILIZING CLOUD FEEDBACK DRAMATICALLY EXPANDS THE HABITABLE ZONE OF TIDALLY LOCKED PLANETS,} The Astrophysical Journal, 771, L45, \dodoi{10.1088/2041-8205/771/2/l45}

\bibitem[{K.~J. Zahnle \& D.~C. Catling(2017)Zahnle \& Catling}]{Zahnle_2017cosmicshoreline}
Zahnle, K.~J., \& Catling, D.~C. 2017, \bibinfo{title}{The Cosmic Shoreline: The Evidence that Escape Determines which Planets Have Atmospheres, and what this May Mean for Proxima Centauri B,} The Astrophysical Journal, 843, 122, \dodoi{10.3847/1538-4357/aa7846}

\bibitem[{S. Zieba {et~al.}(2023)Zieba, Kreidberg, Ducrot, Gillon, Morley, Schaefer, Tamburo, Koll, Lyu, Acuña, Agol, Iyer, Hu, Lincowski, Meadows, Selsis, Bolmont, Mandell, \& Suissa}]{Zieba_2023trappist1c}
Zieba, S., Kreidberg, L., Ducrot, E., {et~al.} 2023, \bibinfo{title}{No thick carbon dioxide atmosphere on the rocky exoplanet TRAPPIST-1 c,} Nature, 620, 746–749, \dodoi{10.1038/s41586-023-06232-z}

\end{thebibliography}
\bibliographystyle{aasjournal}




\end{CJK*}

\end{document}